\documentclass[%
 aip,
 jmp,%
 amsmath,amssymb,
 reprint,%
]{revtex4-1}
\usepackage{amssymb}
\usepackage{amsmath}
\usepackage{cases}
\usepackage{amsthm}
\usepackage{graphicx}
\usepackage{epstopdf}
\usepackage{dcolumn}% Align table columns on decimal point
\usepackage{bm}% bold math

\usepackage{subfigure}

\newtheorem{theorem}{Theorem}
\newtheorem{lemma}[theorem]{Lemma}
\newtheorem{proposition}[theorem]{Proposition}
\newtheorem{corollary}[theorem]{Corollary}
\theoremstyle{definition}

\begin{document}

%\preprint{}

\title{A Full Computation-relevant Topological Dynamics Classification of Elementary Cellular Automata} %Title of paper



\author{Martin Sch\"ule}
\email[]{schuelem@ini.phys.ethz.ch}
\author{Ruedi Stoop}
%\thanks{}
\affiliation{Institute of Neuroinformatics, ETH and University of Zurich, 8057 Zurich, Switzerland}


\begin{abstract}
Cellular automata are both computational and dynamical systems. We give a complete classification of the dynamic behaviour of elementary cellular automata (ECA) in terms of fundamental dynamic system notions such as sensitivity and chaoticity. The ``complex'' ECA emerge to be sensitive, but not chaotic and not eventually weakly periodic. Based on this classification, we conjecture that elementary cellular automata capable of carrying out complex computations, such as needed for Turing-universality, are at the ``edge of chaos''.
\end{abstract}


\maketitle %\maketitle must follow title, authors, abstract and \pacs

\begin{quotation}
In the rich classical history of the theory of computation, models of computation were typically compared to the Turing machine concept, which allows us to characterize their computational power in great detail.\cite{sipser1996introduction,papadimitriou2003computational} If, however, one would like to ascribe ``computational" capacity to processes and systems observed in nature, one is naturally pushed toward using dynamical systemÕs notions as the natural framework, leaving the problem open of how to fit this approach into, or how to link this approach with, Turing computation. A paradigmatic class of systems that comprise in a generic manner both computational and dynamical system aspects are the cellular automata (CA). While CAÕs are defined as a class of discrete dynamical systems, they also serve as a mathematical model of massively parallel computation, a paradigm often observed when ``nature computes''. Remarkably, already very simple rules make CAÕs computationally universal, i.e. capable of carrying out arbitrary computational tasks. By clarifying the dynamical system properties of the most popular and best-studied subclass of CA, the so-called elementary cellular automata (ECA), we will contribute here to a more profound understanding of CA as both computational and dynamical systems. We will fully classify the dynamic behavior of ECA using exclusively topological dynamics attributes such as sensitivity and chaos. Based on this classification, we will finally conjecture that the computationally most complex and biologically relevant ECA are those located at the ``edge of chaos".
\end{quotation}
\section{Introduction}
By definition, CA are discrete dynamical systems acting in a discrete space-time. The state of a CA is specified by the states of the individual cells of the CA, i.e. by the values taken from a finite set of states associated with the sites of a regular, uniform, infinite lattice. The state of a CA then evolves in discrete time steps according to a rule acting synchronously on the states in a finite neighbourhood of each cell. Despite the simplicity of these rules, CA can exhibit strikingly complex dynamical behaviour. A well-known example of a CA with intricate dynamics is the so-called \textit{Game of Life}. CA have also been extensively applied as models for a wide variety of physical and biological processes. 

Obtaining a dynamical system classification of ECA is part of the long-standing problem in CA theory to characterise the ``complexity'' seen inherent in CA behaviour. In a series of influential papers, Wolfram studied the dynamical system and statistical properties of CA and devised a classification scheme.\cite{wolfram1983cellular,wolfram1983statistical,wolfram1984universality} According to this scheme, CA behaviour can be divided into the following classes:\\
\begin{tabular}[t]{ll}
    (W1) & almost all initial configurations lead to the same\\ & fixed point configuration,\\
   (W2) & almost all initial configurations lead to a periodic\\ & configuration,\\
(W3) & almost all initial configurations lead to random\\ & looking behaviour,\\
(W4) & localized structures with complex behaviour\\ & emerge.\\
  \end{tabular}
Wolfram's classification attempt was largely based on simulations of ECA. Since his pioneering work many more classification schemes have been proposed, e.g. by Li et al. \cite{li1990structure} or Culik et al. \cite{culik1988undecidability} It is however still an open problem of CA theory to obtain a completely satisfying, formal classification of CA behaviour.

In this paper, we will put forward a complete topological dynamics classification of ECA. Our approach is based on the symbolic dynamics treatment of CA initiated by the seminal paper of Hedlund.\cite{hedlund1969endomorphisms} The topological dynamics approach allows to use the fundamental notions of dynamics system theory such as sensitivity, chaos, etc. More specifically, the classification is based on a scheme, introduced by Gilman \cite{gilman1987classes} and modified by Kurka \cite{kurka2001languages}, which proposes four classes: Equicontinuous CA, CA with some equicontinuous points, sensitive but not positively expansive CA and positively expansive CA. Each one-dimensional CA belongs to exactly one class, but class membership is generally not decidable.\cite{kurka2001languages} We determine for every ECA, as far as we know for the first time, to which class it belongs. We also (re-)derive further properties such as surjectivity and chaoticity of ECA. Taken together this gives a fairly complete picture of the dynamical system properties of ECA.

The paper is organised as follows. In Sect. \ref{sec2}, we introduce one-dimensional CA and ECA formally. In Sect. \ref{sec3}, we give basic notations and definitions of the topological dynamics approach to CA. In Sect. \ref{sec4}, we introduce a scheme that allows to express ECA rules algebraically. This will prove helpful in Secs. \ref{sec5} and \ref{sec6}, where we will classify ECA in the topologically dynamics sense of Kurka. In Sect. \ref{sec7}, we discuss our results.
\section{Definition of Elementary Cellular Automata}
\label{sec2}
We start with the definitions of the basic concepts underlying the theory of one-dimensional CA. The \textit{configuration} of a one-dimensional CA is given by the double-infinite sequence $x=(x_i)_{i \in \mathbb{Z}}$ with $x_i \in S$ being elements of the finite set of states $S=\{ 0, 1, ...\}$. The configuration space $X$ is the set of all sequences $x$, i.e., $X=S^{\mathbb{Z}}$. The CA map $F$, simply called the CA $F$, is a map $F:X\rightarrow X$ where the \textit{local function} is the map $f: S^{2r+1} \rightarrow S$, $r \geq 1$, with $F(x)_i = f(x_{i-r},...,x_i,...,x_{i+r})$. The integer $r$ is called the \textit{radius} of the CA. The iteration of the map $F$ acting on an initial configuration $x$ generates the \textit{orbit} $x, F(x), F^2(x),...$ of $x$. The orbits of all configurations $x$ are a discrete space-time dynamical system also referred to as CA $F$. Instances of the system can be visualised in so-called \textit{space-time patterns}.

A \textit{spatially periodic} configuration is a configuration which is invariant under translation in space, that is, $x$ is periodic if there is $q > 1$ such that $\sigma^{q}(x)=x$ where $\sigma: X\rightarrow X$ is the \textit{shift map} $\sigma(x)_i=x_{i+1}$. A \textit{temporally periodic} or simply \textit{periodic} configuration $x$ for some CA $F$ is given if $F^n(x)=x$ for some $n > 0$. If $F(x)=x$, $x$ is called a \textit{fixed point}. A configuration $x$ is called \textit{eventually periodic}, if it evolves into a temporally periodic configuration, i.e. if $F^{k+n}(x)=F^k(x)$ for some $k \geq 0$ and $n > 0$. If this holds for any configuration $x$, the corresponding CA is called \textit{eventually periodic}.
 
An elementary cellular automaton (ECA) is an one-dimensional CA with two states and ``nearest neighbourhood coupling'', that is, $S=\{0,1\}$ and $r=1$. There are then $256$ different possible local functions $f: S^3 \rightarrow S$ with $F(x)_i=f({x_{i-1}},{x_i},{x_{i+1}})$. Local functions are also called \textit{rules} and usually given in form of a \textit{rule table}. An example is:
\begin{center}
\begin{tabular}{|c|c|c|c|c|c|c|c|}
111 & 110 & 101 & 100 & 011 & 010 & 001 & 000\\
\hline
0 & 1 & 1 & 0 & 1 & 1 & 1 & 0\\
\end{tabular}
\end{center}
Every ECA rule is, following Wolfram \cite{wolfram1983cellular}, referred to by the sequence of the values of the local function, as given in the rule table, written as a decimal number. In the example above one speaks of ECA rule 110, because $01101110$ written as a decimal number equals $110$.

\section{Topological and Symbolic Dynamics Definitions and Concepts}
\label{sec3}
The framework we use to study the dynamical properties of ECA is given by the symbolic dynamics approach that views the state space $S^{\mathbb{Z}}$ of one-dimensional CA as the Cantor space of symbolic sequences. The topology of the Cantor space is induced by the metric
\begin{equation}
d_C(x,y)=\sum_{i=-\infty}^{+\infty}\frac{\delta(x_i,y_i)}{2^{|i|}}, \nonumber
\label{eqdC}
\end{equation}
where $\delta(x_i,y_i)$ is the discrete metric\\
$\delta(x_i,y_i)=\begin{cases}
1, & x_i\neq y_i \\
0, & x_i = y_i
\end{cases}$.\\
Under this metric the configuration space $S^{\mathbb{Z}}$ is compact, perfect and totally disconnected, i.e., a Cantor space.\cite{kurka2003topological} From now on, the configuration space $S^{\mathbb{Z}}$ endowed with this metric will be referred to as $X$. The ECA functions $F$ are continuous in $X$, hence $(X, F)$ is a (discrete) dynamical system.

Now we introduce some key concepts of the topological dynamics treatment of CA.
A configuration $x$ is an \textit{equicontinuity point} of CA $F$, if $\forall \epsilon > 0, \exists \delta > 0, \forall y \in X:$
\begin{align}
d(x,y) < \delta, \forall n \ge 0: d(F^n(x),F^n(y)) < \epsilon.
\end{align}
If all configurations $x \in X$ are equicontinuity points then the CA is called \textit{equicontinuous}. If there is at least one equicontinuity point, the CA is \textit{almost equicontinuous}.

A CA is \textit{sensitive} (to initial conditions), if $\exists \epsilon > 0, \forall x \in X, \forall \delta > 0, \exists y \in X:$
 \begin{align}
d(x,y) < \delta, \exists n \ge 0: d(F^n(x),F^n(y)) \ge \epsilon.
\end{align}

A CA is \textit{positively expansive}, if
 \begin{align}
\exists \epsilon > 0, \forall x  \neq y \in X, \exists n \ge 0: d(F^n(x),F^n(y)) \ge \epsilon.
\end{align}
Positively expansive CA are sensitive \cite{kurka2003topological}.

If a configuration is an equicontinuity point, its orbit remains arbitrarily close to the orbits of all sufficiently close configurations. If a CA is sensitive, there exists for every configuration at least one configuration arbitrarily close to it such that the orbits of the two configurations will eventually be separated by some constant. Positive expansivity is a stronger form of sensitivity: the orbits of all configurations that differ in some cell will eventually be separated by some constant. The long term behaviour of a sensitive CA can thus only be predicted if the initial configuration is known precisely.

With these concepts, CA as dynamical systems can be classified according to a classification introduced by Gilman \cite{gilman1987classes} and modified by Kurka\cite{kurka2001languages}.
Every one-dimensional CA falls exactly in one of the following classes \cite{kurka2001languages}:\\
\begin{tabular}[t]{l}
(K1) equicontinuous CA,\\
(K2) almost equicontinuous but not equicontinuous CA,\\
(K3) sensitive but not positively expansive CA,\\
(K4) positively expansive CA.\\
\end{tabular}

\begin{figure}[ht]
\centering
\subfigure[\scriptsize{Rule 108}]{
\includegraphics[width=0.22\textwidth]{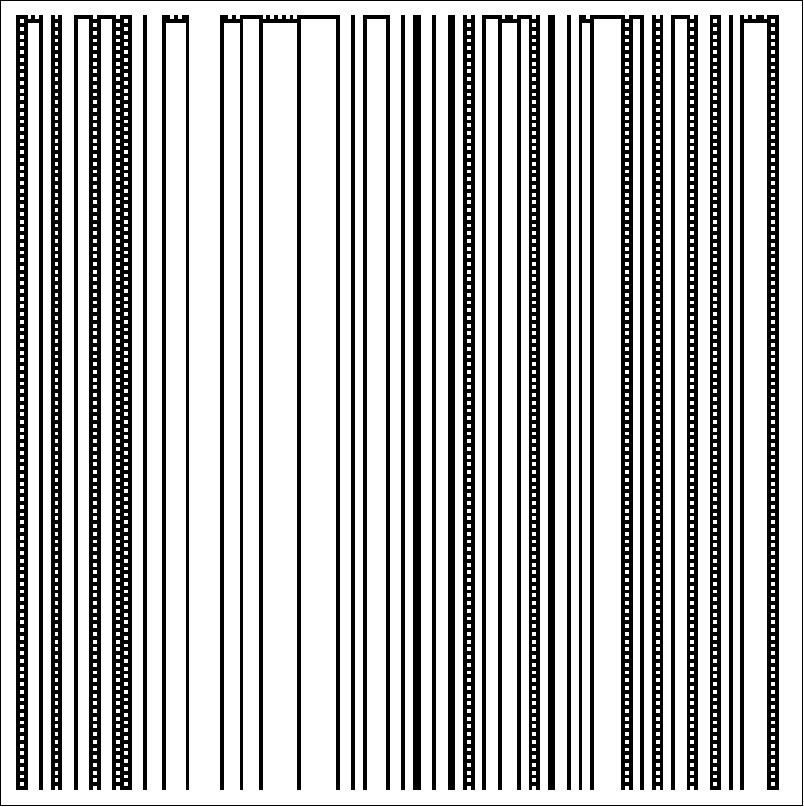}
\label{fig:subfig3}
}
\subfigure[\scriptsize{Rule 73}]{
\includegraphics[width=0.22\textwidth]{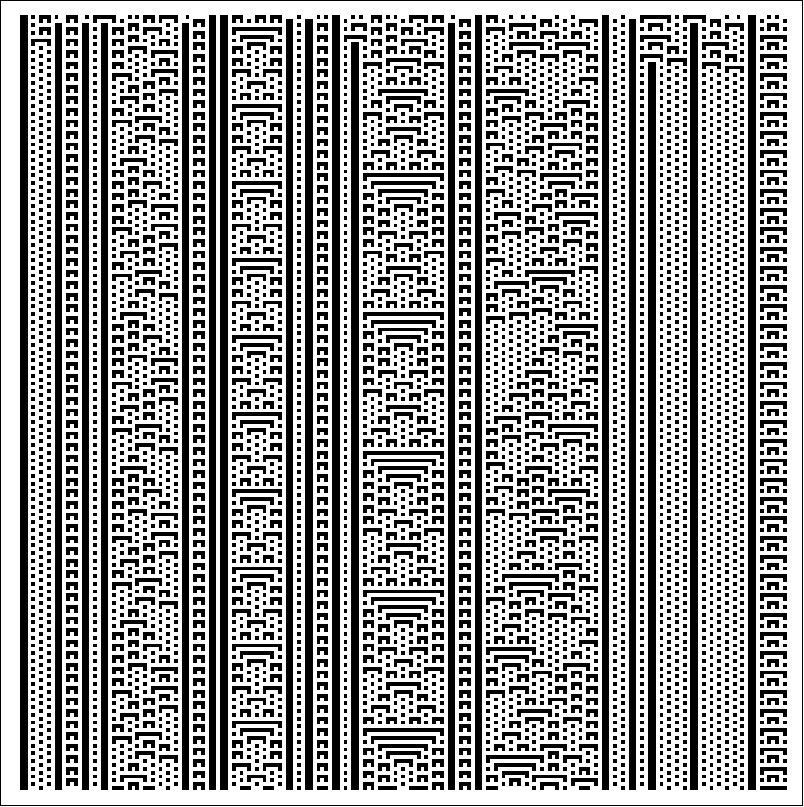}
\label{fig:subfig4}
}
\subfigure[\scriptsize{Rule 110}]{
\includegraphics[width=0.22\textwidth]{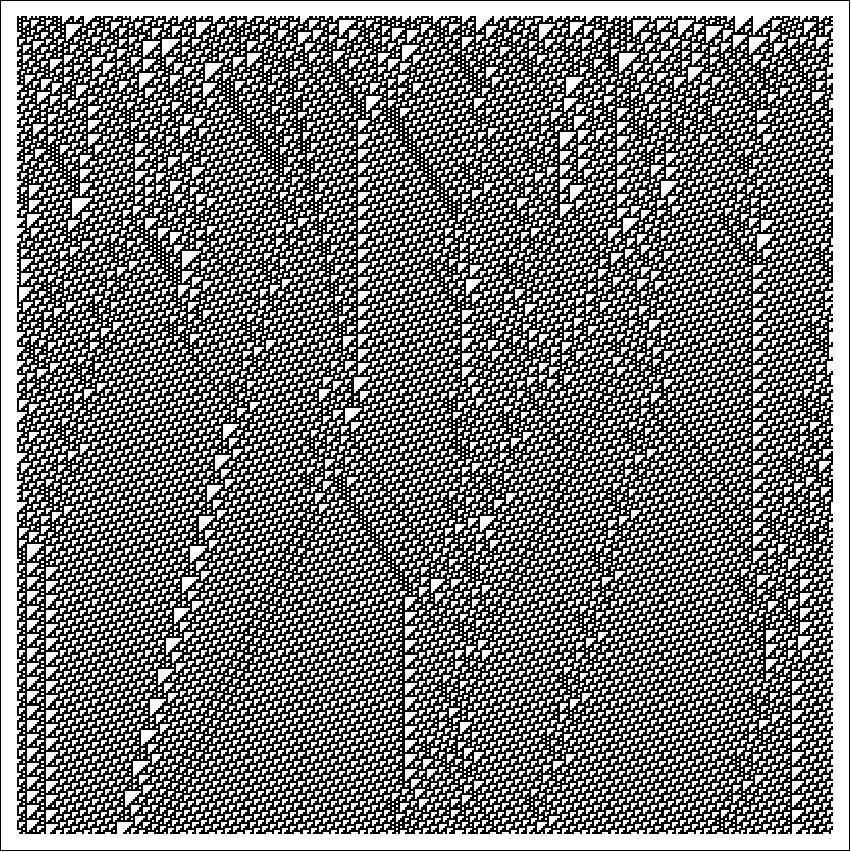}
\label{fig:subfig5}
}
\subfigure[\scriptsize{Rule 90}]{
\includegraphics[width=0.22\textwidth]{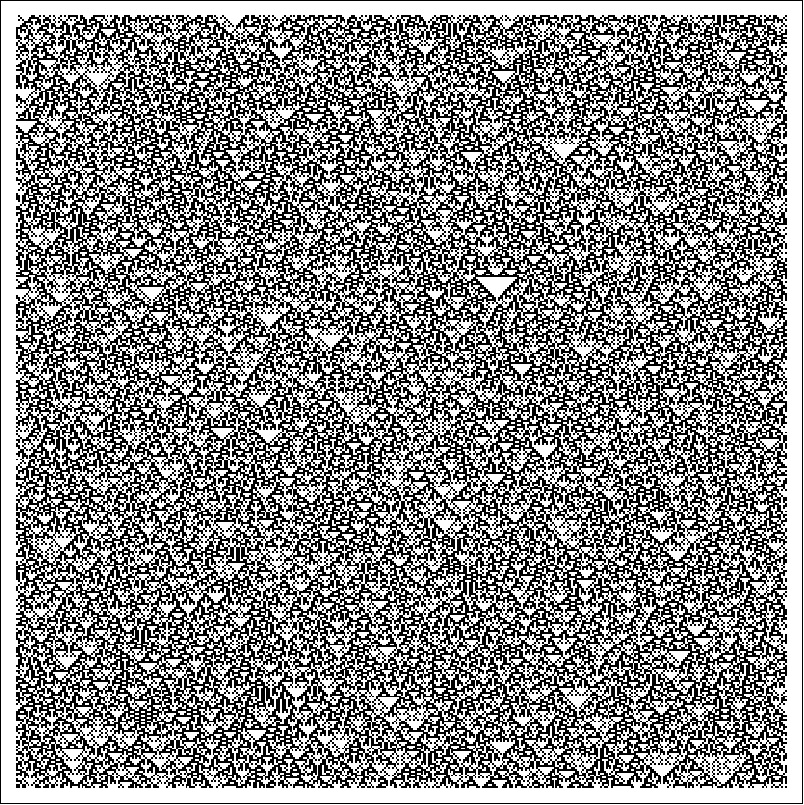}
\label{fig:subfig6}
}
\caption{\scriptsize{Examples of space-time patterns that illustrate the dynamic behaviour of the classes (K1)-(K4): Equicontinuous ECA rule 108 (a), almost equicontinuous but not equicontinuous ECA rule 73 (b), sensitive but not positively expansive ECA rule 110 (c), positively expansive ECA rule 90 (d). Finite arrays of 200 (rule 73 and rule 108) and 400 cells, respectively (rule 110 and rule 90), with periodic boundary conditions are used; black dots code state 1, white dots state 0. Time runs from top to bottom.}}
\label{fig_eca_patterns}
\end{figure}

The typical emergent dynamics of the different classes are illustrated by the space-time patterns of Figure \ref{fig_eca_patterns}.

It has been shown that for one-dimensional CA it is not decidable whether a given CA belongs to class (K1), (K2) or (K3)$\cup$(K4), whereas it is still open whether the class (K4) is decidable.\cite{durand2003undecidability} We will show that it can be determined to which class an ECA belongs.

\section{Algebraic Expressions of Elementary Cellular Automata Rules}
\label{sec4}
Here, we devise an algebraic expression scheme for ECA. The main idea is to derive in a consistent way algebraic expressions for the local ECA rules from a Boolean function form of ECA rules. The algebraic expressions of ECA rules are of use in Secs. V-VI. Algebraic expressions of specific ECA rules have been derived earlier, usually for additive ECA rules.\cite{martin1984algebraic} For example, rule 90 is usually given as ${F(x)_i}={x_{i-1}} + {x_{i+1}} \mod 2$.\cite{kurka2003topological} Other approaches, e.g. by Chua \cite{chua2002nonlinear}, do not yield the same simple polynomial forms as obtained below. The approach taken here was introduced earlier by the present authors\cite{schuele2008global}, where, to the best of our knowledge, for the first time simple, algebraic expressions were given for all ECA rules. Note that Betel and Flocchini used a similar approach in their study on the relationship between Boolean and ``fuzzy'' cellular automata.\cite{betel2009relationship}

The rule tables which define the ECA rules can be regarded as truth tables familiar from propositional logic. Any ECA rule hence corresponds to a \textit{Boolean function}, which can always be expressed as a \textit{disjunctive normal form} (DNF) (or a conjunctive normal form respectively).\cite{mendelson1997introduction} The DNF of a Boolean function is a disjunction of clauses, where a clause is a conjunction of Boolean variables. Any ECA rule can thus be expressed as
\begin{equation}
\bigvee_m \bigwedge_{j=-1}^{1} (\neg)X_{i+j}^m
\end{equation}
where $X_{i+j}$ are Boolean variables associated with the states of the cells in the neighbourhood of an ECA. For example the DNF expression of ECA rule 110 reads to:\\ $(X_{i-1} \wedge  X_{i} \wedge \neg X_{i+1}) \vee (X_{i-1} \wedge \neg X_{i} \wedge X_{i+1}) \vee (\neg X_{i-1} \wedge X_{i} \wedge X_{i+1}) \vee (\neg X_{i-1} \wedge X_{i} \wedge \neg X_{i+1}) \vee (\neg X_{i-1} \wedge   \neg X_{i} \wedge X_{i+1})$.\\ 
The representation of ECA rules in DNF is well known and has e.g. been studied by Wolfram.\cite{wolfram2002new}

We may express now the Boolean operations $(\wedge, \vee, \neg)$ arithmetically as
\begin{align}
x \wedge y &= xy\\
x \vee y &= x + y - xy \nonumber \\
\neg x &= 1 - x. \nonumber
\end{align}
We found it convenient to express the Boolean operations in this way, instead of using the more common modulo-2 operations. This replacement takes the Boolean algebra $(A, \wedge, \vee, \neg, 1, 0)$, with the set $A=\{0,1\}$, into a Boolean ring $(R, +, -, \cdot, 1, 0)$, with the set $R=\{0,1\}$ and the usual arithmetical operations.

Replacing the Boolean operations in the DNF expressions of ECA rules with their arithmetic counterparts yields, for all ECA, Boolean polynomials of the form
\begin{widetext}
{\small
\begin{align}
\alpha_0 + \alpha_1 x_{i-1} + \alpha_2 x_{i} + \alpha_3 x_{i+1} + \alpha_4 x_{i-1}x_{i} + \alpha_5x_{i}x_{i+1} + \alpha_6x_{i-1}x_{i+1}+\alpha_7x_{i-1}x_{i}x_{i+1},
\label{eq6}
\end{align}}
\end{widetext}
with $x_i \in \{0,1\}$ and $\alpha_j \in \mathbb{Z}$. ECA rules are completely determined by the appropriate set of coefficients $\alpha_j$ in expression (\ref{eq6}).

As examples we list here a few algebraic expressions of some interesting ECA rules.\\
\\
Rule 30: ${F(x)_i}=x_{i-1}+x_i +x_{i+1}-2x_{i-1}x_i -x_i x_{i+1}-2x_{i-1}x_{i+1}+2x_{i-1}x_i x_{i+1}$\\
Rule 90: ${F(x)_i}={x_{i-1}} + {x_{i+1}} - 2{x_{i-1}}{x_{i+1}}$\\
Rule 108: ${F(x)_i}={x_{i}} + x_{i-1}{x_{i+1}} - 2{x_{i-1}}x_{i}{x_{i+1}}$\\
Rule 110: ${F(x)_i}={x_{i}} + {x_{i+1}} - {x_{i}}{x_{i+1}} - {x_{i-1}}{x_{i}}{x_{i+1}}$\\
Rule 184: ${F(x)_i}=x_{i-1}-x_{i-1}x_i +x_i x_{i+1}$\\
Rule 232: ${F(x)_i}={x_{i-1}}x_{i} + x_{i}{x_{i+1}} + {x_{i-1}}{x_{i+1}} - 2{x_{i-1}}{x_{i}}{x_{i+1}}$\\

Note how simple, for example, the algebraic expression of the ``complex'' ECA rule 110 is!

It is well-known that the ECA rule space can be partitioned into 88 equivalence classes, because ECA rules are equivalent under the symmetry operations of exchanging left/right and $0/1$ complementation. For the local function $f(x)_i=f({x_{i-1}},{x_i},{x_{i+1}})$ these symmetry operations are given by $T^{left/right}(f(x)_{i})=f({x_{i+1}},{x_i},{x_{i-1}})$ and $T^{0/1}(f(x)_i)=1-f({1-x_{i-1}},{1-x_i},{1-x_{i+1}})$. 

For example, for ECA rule 110 the equivalent rules are:\\
\\
Rule 110: ${F(x)_i}={x_{i}} + {x_{i+1}} - {x_{i}}{x_{i+1}} - {x_{i-1}}{x_{i}}{x_{i+1}}$\\
Rule 137: ${F(x)_i}=1-x_{i-1}-x_i -x_{i+1}+x_{i-1}x_i +2x_i x_{i+1}+x_{i-1}x_{i+1}-x_{i-1} x_i x_{i+1}$\\
Rule 124: ${F(x)_i}=x_{i-1} +x_{i}-x_{i-1} x_{i}-x_{i-1} x_i x_{i+1}$\\
Rule 193: ${F(x)_i}=1-x_{i-1}-x_i -x_{i+1}+2x_{i-1}x_i +x_i x_{i+1}+x_{i-1}x_{i+1}-x_{i-1} x_i x_{i+1}$\\

From now on we will use the lowest decimal ECA rule number present within the group to refer to the whole group. For example, referring to ECA rule 110 implies in this way the four rules $\{110, 137, 124, 193\}$.

Note that the approach developed here can be extended in various ways, for example to one-dimensional CA with state space $\{0,1\}$ with larger neighbourhood, or to two-dimensional CA with state space $\{0,1\}$, etc.

\section{Classification of Elementary Cellular Automata}
\label{sec5}
We will now classify ECA from their topological dynamics properties, that is, according to the scheme introduced by Gilman \cite{gilman1987classes} and modified by Kurka \cite{kurka2001languages}.

First, we need some more symbolic dynamics definitions and notions. A \textit{word} $u$ is a finite symbolic sequence $u=u_0...u_{l-1}$, with $u_i  \in S$, where $S$ is a finite $alphabet$, e.g. in the case of ECA the state set $\{0,1\}$. The length of $u$ is denoted by $l=|u|$. The set of words of $S$ of length $l$ is denoted by $S^l$, the set of all words of $S$ with $l > 0$ is $S^+$. The $cylinder$ $set$ $[u]_{0}$ of $u$ consists of all points $x \in S^{\mathbb{Z}}$ with leading part $u$, i.e. $[u]_{0}=\{x \in S^{\mathbb{Z}}: x_{[0,l)}=u\}$.

\begin{figure}[h!]
  \begin{center}
    \begin{tabular}{c}
\includegraphics[width=0.44\textwidth]{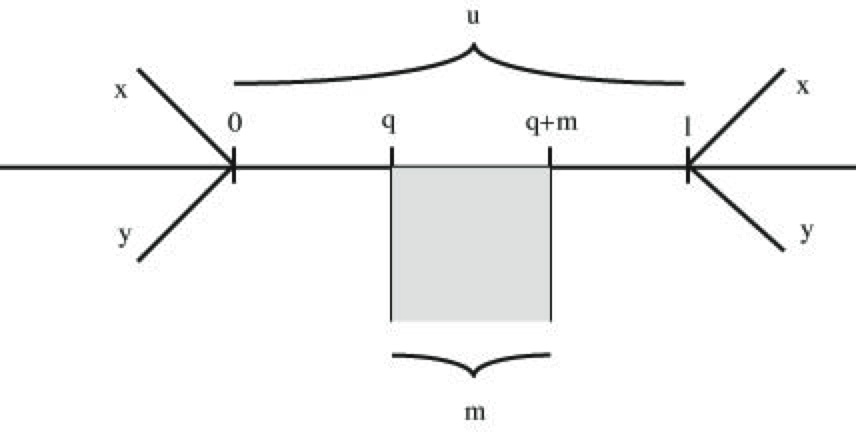}
    \end{tabular}
    \caption[]{\footnotesize A word $u$ of length $|u|=l$ is said to be \textit{blocking}, if it has an interior of size $m$, located from position $q$, that remains unaffected by the states of the cells left and right to the word $u$, at all times.} 
 \label{fig_blocking} 
 \end{center}
\end{figure}

A word $u \in S^{+}$ with $|u| \geq m, m > 0$, is \textit{m-blocking} for a one-dimensional CA $F$, if there exists an offset $q \in [0, |u| - m]$ such that
\begin{center}
$\forall x, y \in [u]_0, \forall n \geq 0, F^n (x)_{[q,q+m)}= F^n (y)_{[q,q+m)}$.
\end{center}
For an illustration of the mathematical definition see Figure \ref{fig_blocking}.

One-dimensional CA, and therefore ECA, are either sensitive or almost equicontinuous. The latter property is equivalent to having a blocking word:

\begin{proposition}[Kurka \cite{kurka2003topological}]
For any one-dimensional CA $F$ with radius $r > 0$ the following conditions are equivalent.
\begin{itemize}
     \item[(1)] $F$ is not sensitive.
     \item[(2)] $F$ has an r-blocking word.
     \item[(3)] $F$ is almost equicontinuous.
     \end{itemize}
     \label{prop2}
\end{proposition}

If a configuration $x$ contains a $m$-blocking word $u$, then the sequence $x_{[q,q+m)}$, i.e. the states of the cells in the segment $[q, q+m)$, are at all times independent of the initial states outside of the blocking word $u$. Hence, the following corollary holds:

\begin{corollary}
For any one-dimensional CA $F$ with radius $r > 0$ the following conditions are equivalent.
\begin{itemize}
     \item[(1)] $F$ has a m-blocking word with $m \geq r$.
     \item[(2)] $F$ has a word $u \in S^{+}$ with $|u| \geq m, m > 0$ and an offset $q \in [0, |u| - m]$ such that
$\forall x \in [u]_0$ the sequence $x_{[q,q+m)}$ is eventually temporally periodic.
     \end{itemize}
     \label{cor1}
\end{corollary}
\begin{proof}
$(1) \Rightarrow (2)$: Denote the sequence $x_{[q,q+m)}$ of a blocking word $u$ that is at all times independent of the initial states outside of $u$ by $v$. The configuration $x=(u)^{\infty}$ is spatially periodic and hence eventually temporally periodic. Because the sequence $v$ is independent of the states of the cells outside of $u$, the sequence $v$ is also eventually temporally periodic.\\
$(2) \Rightarrow (1)$: The condition (2) says that for all $x \in [u]_0$ there is $t \geq 0$ and $p > 0$ such that $F^{t+p}(x)_{[q,q+m)}=F^{t}(y)_{[q,q+m)}$. Thus, for all $x,y \in [u]_0$ and all $n \geq 0$ the sequence $F^n (x)_{[q,q+m)}=F^n (y)_{[q,q+m)}$ must be independent of the initial states outside of $u$, hence the word $u$ is $m$-blocking.
\end{proof}

We will now systematically search for blocking words. We know by proposition \ref{prop2} that whenever a blocking word can be found, the corresponding ECA is almost equicontinuous. By corollary \ref{cor1}, we know that this corresponds to finding a word $u$ that contains a sequence that is eventually temporally periodic, independent of the initial states outside of $u$. As it turns out, we can thereby effectively determine all almost equicontinuous ECA, because any almost equicontinuous ECA corresponds to a blocking word $u$ for which the length $l=|u|$ is bounded.

\begin{proposition}
Each almost equicontinuous ECA has at least one blocking word of length $l \leq 4$.
\label{prop4}
\end{proposition}

\begin{proof}
In the following, we look for blocking words, starting with the smallest possible length $l=1$ and then successively for words of greater length (for a visualisation of the definition of a blocking word see again Figure \ref{fig_blocking}). If a blocking word can be found, one or several almost equicontinuous ECA rules will satisfy the blocking conditions. The ECA rules are specified by a rule table which we denote by $(t_0, t_1, t_2, t_3, t_4, t_5, t_6, t_7)$. For example, ECA rule 110 is given by the table $(0,1,1,0,1,1,1,0)$. If an entry in the rule table is left unspecified, the entry can take on either of the two values 0 or 1, e.g. the table $(0,1,1,0,1,1,1, t_7)$ refers to the two ECA rules 110 and 111. If a blocking word can be found, we put the ECA rule table admitted by the blocking conditions in a list. A blocking word $u$ and the admitted rule table is denoted by $t(u,p)=(t_0, t_1, t_2, t_3, t_4, t_5, t_6, t_7)$, where $p$ is the period with which the eventually periodic sequence in the word $u$ (i.e. the sequence $x_{[q,q+m)}$ referred to in corollary \ref{cor1}) is repeated. For example, $t(00,1)=(t_0, t_1,t_2, 0, t_4, t_5, 0, 0)$ refers to the blocking word $00$ of period $p=1$ that corresponds to $2^5=32$ ECA rules, as denoted by the rule table. If a newly found blocking word admits ECA rules generated by a rule table obtained by a blocking word already in the list (hence of smaller length), the word and the rule table admitted by it is not listed. We also do not list blocking words, and the rule tables admitted by them, if they correspond to ECA rules equivalent to ECA rules admitted by a blocking word already in the list.

Let us further assume the following notation: The variable $c_i$ always denotes the states of cells $i$ of a blocking word $u$ that are at all times independent of the initial states of the cells outside of the blocking word $u$. The variable $x_i$ on the other hand denotes the states of cells $i$ that are in principle influenceable by the initial states of the cells outside of $u$. The state $x_i$ of such a cell $i$ is left undetermined, i.e. the value can either be 0 or 1. If it is known for configurations $x, y \in [u]_0$ that the states $x_i$ and $y_i$ of some cell $i$ differ, we write $\bar{x}_i$. For example, the ``scenario'' $\begin{array}{ccc} \bar{x}_{-1} & c_0 & \bar{x}_{1} \\ x_{-1} & c_0 & x_{1} \end{array}$ refers to two configurations $x, y \in [u]_0$ that share the blocking word $u=c_0$ of length $l=1$ that is repeated with period $p=1$. At the boundaries of the blocking word $u$, here at the cells $i=-1$ and $i=1$, we can assume that the configurations $x$ and $y$ differ, which is denoted by $\bar{x}_{-i}$ and $\bar{x}_i$, whereas in the next time step this may not necessarily be the case anymore (at the cells $i=-1$ and $i=1$).

The proof has two parts. In part A, we determine all blocking words of length $l \leq 4$. In part B, we show that for any blocking word $u$ of length $l > 4$ there is a corresponding blocking word of length $l \leq 4$.

Part A: Let us look at the cases (a) $l=1$, (b) $l=2$, (c) $l=3$ and (d) $l=4$, where $l$, as said, denotes the length of a blocking word $u$.

(a) With $l=1$, the following scenarios are possible: (1) $\begin{array}{ccc} \bar{x}_{-1} & c_0 & \bar{x}_{1} \\ x_{-1} & c_0 & x_{1} \end{array}$, (2) $\begin{array}{ccc} \bar{x}_{-1} & c_0 & \bar{x}_{1} \\ c_{0} & c_0 & x_{1} \end{array}$, (3) $\begin{array}{ccc} \bar{x}_{-1} & c_0 & \bar{x}_{1} \\ x_{-1} & c_0 & c_{0} \end{array}$, (4) $\begin{array}{ccc} \bar{x}_{-1} & c_0 & \bar{x}_{1} \\ c_{0} & c_0 & c_{0} \end{array}$, (5) $\begin{array}{ccc} \bar{x}_{-1} & c_0 & \bar{x}_{1} \\ x_{-1} & c'_0 & x_{1} \end{array}$, (6) $\begin{array}{ccc} \bar{x}_{-1} & c_0 & \bar{x}_{1} \\ c'_{-1} & c'_0 & x_{1} \end{array}$,\\ (7) $\begin{array}{ccc} \bar{x}_{-1} & c_0 & \bar{x}_{1} \\ x_{-1} & c'_0 & c'_{1} \end{array}$ and (8) $\begin{array}{ccc} \bar{x}_{-1} & c_0 & \bar{x}_{1} \\ c'_{-1} & c'_0 & c'_{1} \end{array}$, where at least for one $i$, $c'_i \neq c_i$. Note that there are further scenarios possible that however do not yield further valid rule tables and are not listed here. Scenario (1) yields the rule table $t(1,1)=(1,1,t_2, t_3, 1, 1, t_6, t_7)$ (and the table $t(0,1)=(t_0, t_1, 0, 0, t_4, t_5, 0, 0)$, but as said, tables that yield ECA rules equivalent to already obtained rules are not listed). Scenarios (2), (3) and (4) do not admit rule tables that yield ECA rules not already listed. For scenario (5), two cases have to be further distinguished: $\begin{array}{ccc} \bar{x}_{-1} & c_0 & \bar{x}_{1} \\ x_{-1} & c'_0 & x_{1} \\  & c'_0 &  \end{array}$ and $\begin{array}{ccc} \bar{x}_{-1} & c_0 & \bar{x}_{1} \\ x_{-1} & c'_0 & x_{1} \\  & c_0 &  \end{array}$, where $c'_0 \neq c_0$. The first case does not lead to new ECA rules. The second case yields the rule table $t(1,2)=(0, 0, 1, 1, 0, 0, 1, 1)$. Scenarios (6) and (7) yield rule tables already listed. Scenario (8) yields $t(1,2)=(0, 0, 0, 0, 0, 0, 0, 1)$.

(b) For $l=2$, we deal with essentially the same scenarios as in case (a). For example, in analogy to the scenario (1) $\begin{array}{ccc} \bar{x}_{-1} & c_0 & \bar{x}_{1} \\ x_{-1} & c_0 & x_{1} \end{array}$ of case (a), we have the scenario $\begin{array}{cccc} \bar{x}_{-1} & c_0 & c_{1} & \bar{x}_{2} \\ x_{-1} & c_0 & c_{1} & x_{2} \end{array}$. However, for reasons of space, we cannot list all possible scenarios and from now on only list the scenarios that yield blocking words that admit rule tables not yet obtained. These are: (1) $\begin{array}{cccc} \bar{x}_{-1} & c_0 & c_{1} & \bar{x}_{2} \\ x_{-1} & c_0 & c_{1} & x_{2} \end{array}$ and (2) $\begin{array}{cccc} \bar{x}_{-1} & c_0 & c_{1} & \bar{x}_{2} \\ x_{-1} & c'_0 & c'_{1} & x_{2} \end{array}$, where $c'_i \neq c_i$. Scenarios (1) and (2) yield, as can easily be checked, the following blocking words and rule tables: $t(00,1)=(t_0, t_1,t_2, 0, t_4, t_5, 0, 0)$, $t(01,1)=(t_0, t_1,0, t_3, 1, 1, 0, t_7)$, $t(10,1)=(t_0,1,0, 0, t_4, 1, t_6, t_7)$ and $t(00,2)=(0,0,t_2, 1, 0, t_5, 1, 1)$, $t(10,2)=(t_0,0,1, 1, 0, 0, 1, t_7)$.

(c) For $l=3$, the scenarios that yield rule tables not listed above are: (1) $\begin{array}{ccccc} \bar{x}_{-1} & c_0 & c_{1} &  c_{2} &\bar{x}_{3} \\ x_{-1} & c_0 & c_{1} & c_{2} & x_{3} \end{array}$ and (2) $\begin{array}{ccccc} \bar{x}_{-1} & c_0 & c_{1} &  c_{2} &\bar{x}_{3} \\ x_{-1} & \bar{x}_0 & c'_{1} & \bar{x}_{2} & x_{3} \\ & c_0 & c_{1} & c_{2} & \end{array}$, where $c'_1 \neq c_1$. Scenario (1) yields the blocking words and rule tables $t(010,1)=(t_0,t_1,0, 0, t_4, 1, 0, t_7)$ and $t(101,1)=(t_0,1,0, t_3, 1, 1, t_6, t_7)$. Scenario (2) yields $t(000,2)=(0,0,t_2, 0, 0, 0, 0, 1)$. Note that for example the scenario\\ $\begin{array}{ccccc} \bar{x}_{-1} & c_0 & c_{1} &  c_{2} &\bar{x}_{3} \\ x_{-1} & \bar{x}_0 & c_{1} & \bar{x}_{2} & x_{3} \\ & c_0 & c_{1} & c_{2} & \end{array}$\\ does not yield new rule tables.

(d) For $l=4$, the only scenario that leads to a blocking word corresponding to a rule table not yet listed is $\begin{array}{cccccc} \bar{x}_{-1} & c_0 & c_{1} &  c_{2} & c_{3} &\bar{x}_{4} \\ x_{-1} & c_0 & c_{1} & c_{2} & c_{3} & x_{4} \end{array}$, yielding the rule table $t(0110,1)=(t_0, 1, 0, 0, 1, t_5, 0, t_7)$. Note again that e.g. the scenario $\begin{array}{cccccc} \bar{x}_{-1} & c_0 & c_{1} &  c_{2} &c_{3} &\bar{x}_{4} \\ x_{-1} & \bar{x}_0 & c'_{1} & c'_{2} & \bar{x}_{3} & x_{4} \\ & c_0 & c_{1} & c_{2} & c_{3} & \end{array}$, where at least for one $i$ $c'_i \neq c_i$, does not yield new rule tables.

With this we conclude Part A. Let us list the blocking words and the rule tables admitted by them that we have found so far:
\begin{center}
\begin{tabular}{|l|}
\hline
$t(0,1)=(t_0, t_1, 0, 0, t_4, t_5, 0, 0)$\\
$t(1,2)=(0, 0, 1, 1, 0, 0, 1, 1)$\\
$t(1,2)=(0, 0, 0, 0, 0, 0, 0, 1)$\\
$t(00,1)=(t_0, t_1,t_2, 0, t_4, t_5, 0, 0)$\\
$t(01,1)=(t_0, t_1,0, t_3, 1, 1, 0, t_7)$\\
$t(10,1)=(t_0,1,0, 0, t_4, 1, t_6, t_7)$\\
$t(00,2)=(0,0,t_2, 1, 0, t_5, 1, 1)$\\
$t(01,2)=(t_0,0,1, 1, 0, 0, 1, t_7)$\\
$t(010,1)=(t_0,t_1,0, 0, t_4, 1, 0, t_7)$\\
$t(101,1)=(t_0,1,0, t_3, 1, 1, t_6, t_7)$\\
$t(000,2)=(0,0,t_2, 0, 0, 0, 0, 1)$\\
$t(0110,1)=(t_0, 1, 0, 0, 1, t_5, 0, t_7)$\\
\hline
\end{tabular}
\end{center}

Part B: In the general case, i.e. for $l > 4$, we can conclude in analogy to the cases already considered, i.e. the cases with $l\leq 4$, that the following scenarios could possibly lead to new blocking words:\\
\begin{widetext}
(1) $\left( \begin{array}{ccccccc} \bar{x}_{-1} & c_0 & c_{1} & ... & c_{l-2} & c_{l-1} & \bar{x}_{l} \\ x_{-1} & c_0 & c_{1} & ... & c_{l-2} & c_{l-1} & x_{l} \end{array} \right)$, 
(2) $\left( \begin{array}{ccccccc} \bar{x}_{-1} & c_0 & c_{1} & ... & c_{l-2} & c_{l-1} & \bar{x}_{l} \\ c_{-1} & c_0 & c_{1} & ... & c_{l-2} & c_{l-1} & c_{l} \end{array} \right)$,\\
(3) $\left( \begin{array}{ccccccccccc} \bar{x}_{-1} & c_0 & c_{1} & ... &&&& ... & c_{l-2} & c_{l-1} & \bar{x}_{l} \\ &&&&&...&&&&&\\  &&& \bar{x}_{q-1} & c_{q} & ... & c_{q+m-1} & \bar{x}_{q+m} &&&\\ &&& x_{q-1} & c_{q} & ... & c_{q+m-1} & x_{q+m} &&& \end{array} \right)$,\\
(4) $\left( \begin{array}{ccccccccccc} \bar{x}_{-1} & c_0 & c_{1} & ... &&&& ... & c_{l-2} & c_{l-1} & \bar{x}_{l} \\ &&&&&...&&&&&\\  &&& \bar{x}_{q-1} & c_{q} & ... & c_{q+m-1} & \bar{x}_{q+m} &&&\\ &&& c_{q-1} & c_{q} & ... & c_{q+m-1} & c_{q+m} &&& \end{array} \right)$,
(5) $\left( \begin{array}{ccccccc} \bar{x}_{-1} & c_0 & c_{1} & ... & c_{l-2} & c_{l-1} & \bar{x}_{l} \\ x_{-1} & c'_0 & c'_{1} & ... & c'_{l-2} & c'_{l-1} & x_{l} \end{array} \right)$,\\
(6) $\left( \begin{array}{ccccccc} \bar{x}_{-1} & c_0 & c_{1} & ... & c_{l-2} & c_{l-1} & \bar{x}_{l} \\ c'_{-1} & c'_0 & c'_{1} & ... & c'_{l-2} & c'_{l-1} & c'_{l} \end{array} \right)$,
(7)  $\left( \begin{array}{ccccccccccc} \bar{x}_{-1} & c_0 & c_{1} & ... &&&& ... & c_{l-2} & c_{l-1} & \bar{x}_{l} \\ &&&&&...&&&&&\\  &&& \bar{x}_{q-1} & c_{q} & ... & c_{q+m-1} & \bar{x}_{q+m} &&&\\ &&& x_{q-1} & c'_{q} & ... & c'_{q+m-1} & x_{q+m} &&& \end{array} \right)$,\\
(8) $\left( \begin{array}{ccccccccccc} \bar{x}_{-1} & c_0 & c_{1} & ... &&&& ... & c_{l-2} & c_{l-1} & \bar{x}_{l} \\ &&&&&...&&&&&\\  && \bar{x}_{q-1} & c_{q} & c_{q+1} & ... & c_{q+m-2}& c_{q+m-1} & \bar{x}_{q+m} &&\\ &&& \bar{x}_{q} & c'_{q+1} & ... & c'_{q+m-2} & \bar{x}_{q+m-1} &&&\\ &&& c_{q} & c_{q+1} & ... & c_{q+m-2} & c_{q+m-1} &&& \end{array} \right)$, with $m \geq 1$ and where at least for one $i$, $c'_i \neq c_i$.
\end{widetext}

Case (1) yields blocking words already listed, because for $l>4$ the conditions to be satisfied in order to obtain a blocking word $u$ are entailed in the conditions to obtain a blocking word $u$ with $l\leq 4$. The same reasoning applies to cases (2), (3) and (4). The basic reason that such a reduction is possible is due to the fact that the conditions to be satisfied in order to obtain a blocking word depend on the values of the boundary cells, here the values $\bar{x}_{-1}$ and $\bar{x}_{l}$ (respectively the values $\bar{x}_{q-1}$ and $\bar{x}_{q+m}$ in cases (3) and (4)), but not on the values of the cells to the left (of $i=-1$) and right (of $i=l$) of the boundary cells, as can be checked with the scenarios treated in Part A.

Let us then look closer at case (5). We will show that if there is a blocking word $c_1c_2...c_{l-2}c_{l-1}$, the word is repeated with period $p=2$, because if the word is blocking, the word at the next time step (in case (5)) must be $\bar{c}_1 \bar{c}_2...\bar{c}_{l-2} \bar{c}_{l-1}$. The bar signifies that the state $c_i$ of the cell $i$ must change, i.e. $\bar{c}_i=(1-c_i)$. Without loss of generality, we can consider only the $2^4$ boundary conditions for blocking words at successive time steps. That is, given the word $c_{1}c_{2}...c_{l-2}c_{l-1}$, we consider at the next time-step all the $(2^4-2)$ possible cases: $c_{1}c_{2}...c_{l-2}\bar{c}_{l-1}$, $c_{1}c_{2}...\bar{c}_{l-2}c_{l-1}$, etc., excluding the two cases $c_{1}c_{2}...c_{l-2}c_{l-1}$ and $\bar{c}_1\bar{c}_2...\bar{c}_{l-2}\bar{c}_{l-1}$. It suffices to consider the case $c_{1}c_{2}...c_{l-2}\bar{c}_{l-1}$. The other cases can be dealt with analogously. The temporal evolution of the ECA generates in this case the following scheme:
\begin{center}
$\begin{array}{ccccccc} \bar{x}_{-1} & c_0 & c_{1} & ... & c_{l-2} & c_{l-1} & \bar{x}_{l} \\ x_{-1} & c_0 & c_{1} & ... & c_{l-2} & \bar{c}_{l-1} & x_{l} \\  x_{-1} & c_0 & c_{1} & ... & \bar{c}_{l-2} & c^2_{l-1} & x_{l} \\  &  &  & ... &  &  & \\  x_{-1} & c_0 & \bar{c}_{1} & ... & c^{l-2}_{l-2} & c^{l-2}_{l-1} & x_{l} \\  x_{-1} & \bar{c}_0 & c^{l-1}_{1} & ... & c^{l-1}_{l-2} & c^{l-1}_{l-1} & x_{l} \end{array}$
\end{center}
The superscript denotes the time-step $n$. The third, fifth and sixth line are due to the fact that if the state of e.g. the cell $l-2$ at time step $n=2$ did not change, one would obtain a blocking word of shorter length ($l-1$). By checking all $2^{4}$ possible values for the boundary states of the initial word $c_{1}c_{2}...c_{l-2}c_{l-1}$ it can be shown that the above scheme cannot be satisfied. Thus, any initial word $c_{1}c_{2}...c_{l-2}c_{l-1}$ evolves in the next time step into either the word $c_{1}c_{2}...c_{l-2}c_{l-1}$ or the word $\bar{c}_1\bar{c}_2...\bar{c}_{l-2}\bar{c}_{l-1}$. In the first case, a blocking word of period $p=1$ is found, in the second case, i.e. for $p=2$, one can find a blocking word of length $l=2$, as can easily be shown.

The case (6) can be reduced to the case already treated under (a (8)) in Part A, the case (7) to the case (5) and the case (8) again to the case treated under (c (2)) (or the example in (d) respectively) in Part A. 

With this we conclude our analysis. In Part A, we have identified all blocking words of length $l \leq 4$. For  $l \geq 2$, we omitted, for reasons of space, the presentation of the cases that do not lead to blocking words or to blocking words already identified. In Part B, we have concluded from the cases for $l \leq 4$ on the general form of the scenarios that could possibly lead to blocking words for $l > 4$. These general scenarios could then be reduced to the scenarios obtained for $l \leq 4$. One case (case (5)) required a separate treatment and was analysed by means of an example. 

To arrive at a complete list of blocking words for $l \leq 4$ and to exclude additional blocking words for $l>4$, great care and efforts have been invested. We have tested the completeness of the list also by extensively sampling the space of initial configurations for ECA, which yielded no additional blocking words. One may also check the correctness and completeness of the cases investigated in our analysis by hand with the help of a computer, running a program that follows the lines of the proof above. Alternatively, to demonstrate the impossibility of additional blocking words, the systems of equations generated from the conditions for blocking words and the algebraic expressions of ECA rules could be used, systematically evaluated for each single case.

\end{proof}

The proof of proposition \ref{prop4} allows to give, for ECA, a stronger version of proposition \ref{prop2}. Let us call a word $u$ of length $l$ $invariant$ for an ECA $F$, if for all $x \in [u]_{0}$, there is a $p > 0$ such that $F^{p} (x)_{[0,l)}= x_{[0,l)}$.
\begin{corollary}
An ECA $F$ is almost equicontinuous if and only if $F$ has an $invariant$ word.
 \label{cor3_1}
\end{corollary}
\begin{proof}
See the proof of proposition \ref{prop4}.
\end{proof}

Proposition \ref{prop4} (or corollary \ref{cor3_1} respectively) allows us to determine for each ECA rule whether it is almost equicontinuous or not. It is almost equicontinous if there is an associated blocking word on the list composed of the invariant words of shortest length. Below we provide this list together with the corresponding almost equicontinuous ECA rules.
\begin{corollary}
Invariant words of period $p=1$ and corresponding ECA rules:\\
$0$:
0, 4, 8, 12, 72, 76, 128, 132, 136, 140, 200, 204.\\ 
$00$:
32, 36, 40, 44, 104, 108, 160, 164, 168, 172, 232.\\
$01$:
13, 28, 29, 77, 156.\\
$10$:
78.\\
$010$:
5.\\
$101$:
94.\\
$0110$:
73.\\
Invariant words of period $p = 2$ and corresponding ECA rules:\\
$1$:
1.\\
$0$:
51.\\
$00$:
19, 23.\\
$01$:
50, 178.\\
$000$:
33.
\label{corollary6}
\end{corollary}
Conversely, we now also know the sensitive ECA rules.
 \begin{proposition}
The following rules are sensitive:\\
2, 3, 6, 7, 9, 10, 11, 14, 15, 18, 22, 24, 25, 26, 27, 30, 34, 35, 37, 38, 41, 42, 43, 45, 46, 54, 56, 57, 58, 60, 62, 74, 90, 105, 106, 110, 122, 126, 130, 134, 138, 142, 146, 150, 152, 154, 162, 170, 184.
\label{prop6}
\end{proposition}
\begin{proof}
Follows from Proposition \ref{prop2}, \ref{prop4} and corollary \ref{corollary6}.
\end{proof}
The class of sensitive ECA is large, because in the Cantor space left- or right-shifting rules are sensitive. We will later return to this point.

From the almost equicontinuous ECA rules, we can further specify the \textit{equicontinuous} ones. We use the following lemma.
\begin{lemma}[Kurka \cite{kurka2003topological}]
A one-dimensional almost equicontinuous CA $F$ is equicontinuous if and only if:
\begin{itemize}
     \item[(1)] There exists a preperiod $m \geq 0$ and a period $p > 0$, such that $F^{m+p}=F^m$.
     \end{itemize}
     It is almost equicontinuous but not equicontinuous if and only if:
     \begin{itemize}
     \item[(2)] There is at least one point $x \in X$ for which the almost equicontinuous CA $F$ is not equicontinuous.
     \end{itemize}
     \label{lemma8}
\end{lemma}
\begin{proposition}
The following rules are equicontinuous:\\
0, 1, 4, 5, 8, 12, 19, 29, 36, 51, 72, 76, 108, 200, 204.
\end{proposition}
\begin{proof}
The proof is by showing that condition $(1)$ of Lemma \ref{lemma8} holds. We only give an example for a specific ECA rule.

Rule 72 is equicontinuous with preperiod $m=2$ and period $p=1$, because, by using the algebraic expression for the local function, we obtain\\
${F(x)_i}=x_{i-1}x_i +x_{i}x_{i+1}-2x_{i-1} x_i x_{i+1}$\\
${F^2(x)_i}=x_{i-1}x_{i}-x_{i-2}x_{i-1}x_{i}+ x_{i}x_{i+1}-2x_{i-1}x_{i}x_{i+1}+x_{i-2}x_{i-1}x_{i}x_{i+1}-x_{i}x_{i+1}x_{i+2}+x_{i-1}x_{i}x_{i+1}x_{i+2}$\\
${F^3(x)_i}=x_{i-1}x_{i}-x_{i-2}x_{i-1}x_{i}+ x_{i}x_{i+1}-2x_{i-1}x_{i}x_{i+1}+x_{i-2}x_{i-1}x_{i}x_{i+1}-x_{i}x_{i+1}x_{i+2}+x_{i-1}x_{i}x_{i+1}x_{i+2}$.\\
Hence, ${F^3(x)_i}={F^2(x)_i}$, $\forall i \in \mathbb{Z}$. Thus, $F^3=F^2$.
\end{proof}
\begin{proposition}
The following rules are almost equicontinuous but not equicontinuous:\\
13, 23, 28, 32, 33, 40, 44, 50, 73, 77, 78, 94, 104, 128, 132, 136, 140, 156, 160, 164, 168, 172, 178, 232.
\end{proposition}
\begin{proof}
The proof is by showing that condition $(2)$ of Lemma \ref{lemma8} holds. We only give an example for a specific ECA rule.

ECA rule 104 is almost equicontinuous but not equicontinuous, because $(10)^{\infty}$ is not an equicontinuous point.\\
Assume the configuration $x=(10)^{\infty}$ and an integer $q > 0$ such that 
\begin{align}
\forall y \in X, (x_{[-q,q]}=y_{[-q,q]}) \Rightarrow (d(x,y) < 2^{-q}). \nonumber
\end{align}
Assume that $y$ differs from $x$ at cells $(-q-1)$ and $(q+1)$, that is $y_{-q-1}=1-x_{-q-1}$ and $y_{q+1}=1-x_{q+1}$. Then, as can easily be shown by using the algebraic expression of ECA rule 104, 
\begin{align}
d(F^n(x),F^n(y)) > 2^{-(q-n)} \nonumber
\end{align}
for all $n \leq q$. Hence, ECA 104 is not equicontinuous at the point $x=(10)^{\infty}$.
\end{proof}

From the sensitive ECA, we can distinguish further the \textit{positively expansive} ECA. 

First, we need the definition of \textit{permutivity} for ECA.\cite{kurka2008topological} An ECA $F$ is \textit{left-permutive} if $(\forall u \in S^2), (\forall b \in S), (\exists ! a \in S)$: $f(au)=b$. It is \textit{right-permutive} if $(\forall u \in S^2), (\forall b \in S), (\exists ! a \in S)$: $f(ua)=b$. The ECA $F$ is \textit{permutive} if it is either left-permutive or right-permutive.

We will use the following lemma:

\begin{lemma}[Kurka \cite{kurka2003topological}]
A one-dimensional CA $F$ is positively expansive if the following condition holds.
\begin{itemize}
     \item[(1)] The CA is both left- and right-permutive.
     \end{itemize}
     A one-dimensional sensitive CA $F$ is not positively expansive if and only if the following condition holds.
     \begin{itemize}
     \item[(2)] There is no $\epsilon > 0$ such that for all $x \neq y \in X$ there is $n \ge 0$ with $d(F^n(x),F^n(y)) \ge \epsilon$.
     \end{itemize}
     \label{lemma9}
\end{lemma}
\begin{proposition}
The following ECA rules are sensitive but not positively expansive:\\
2, 3, 6, 7, 9, 10, 11, 14, 15, 18, 22, 24, 25, 26, 27, 30, 34, 35, 37, 38, 41, 42, 43, 45, 46, 54, 56, 57, 58, 60, 62, 74, 106, 110, 122, 126, 130, 134, 138, 142, 146, 152, 154, 162, 170, 184.
\label{prop13}
\end{proposition}
\begin{proof}
The proof is by showing that condition $(2)$ of Lemma \ref{lemma9} holds. We only provide the example of a specific ECA rule.\\
ECA rule 110 is sensitive but not positively expansive. Assume the expansivity constant $\epsilon=2^{-m}$, then 
\begin{align}
\forall x \neq y \in X \Rightarrow \exists n \ge 0, F^n(x)_{[-m,m]}\neq F^n(y)_{[-m,m]}
\label{proofprop11}
\end{align}
must hold. Assume the configuration $x=(00110111110001)^{\infty}$ and an integer $q > 0$ such that $14q > m$. Then, for a configuration $y \in X$ that differs from $x$ at the cells $14q, 14q+1, 14q+2$, (\ref{proofprop11}) does not hold.
\end{proof}

\begin{table*}
\caption{Topological dynamics classification of ECA rules.}
\begin{tabular}{|c|c||c|c|}
\hline
\multicolumn{2}{|c||}{almost equicontinuous} & \multicolumn{2}{|c|}{sensitive}\\
\hline
equicontinuous & & & positively expansive\\
\hline
0, 1, 4, 5, 8, & 13, 23, 28, 32, & 2, 3, 6, 7, 9, & 90, 105, 150\\
12, 19, 29, 36, & 33, 40, 44, 50, 73, & 10, 11, 14, 15, &\\
51, 72, 76, 108, & 77, 78, 94, 104, & 18, 22, 24, 25, &\\
200, 204 & 128, 132, 136, 140, & 26, 27, 30, 34, &\\
& 156, 160, 164, 168, & 35, 37, 38, 41, &\\
& 172, 178, 232 &  42, 43, 45, 46, &\\
& & 54, 56, 57, 58, &\\
& & 60, 62, 74, 106, &\\
& & 110, 122, 126, 130, &\\
& & 134, 138, 142, 146, &\\
& & 152, 154, 162, 170, 184 &\\
\hline
\end{tabular}
\label{table_1}
\end{table*}

\begin{proposition}
The following ECA rules are positively expansive:\\
90, 105, 150. 
\label{prop14}
\end{proposition}
\begin{proof}
For ECA rules 90, 105 and 150 condition $(1)$ of Lemma \ref{lemma9} holds.
\end{proof}
For ECA left- and right-permutivity is equivalent to positive expansivity.
\begin{proposition}
ECA are positively expansive if and only if they are both left- and right-permutive.
\label{prop15}
\end{proposition}
\begin{proof}
Follows from Proposition \ref{prop6}, \ref{prop13} and \ref{prop14}.
\end{proof}
Note that Proposition \ref{prop15} does not hold generally for one-dimensional CA.\cite{kurka2003topological}

We summarize the findings of this section in Table \ref{table_1}, which shows all ECA rules according to whether they have the property of equicontinuity, almost equicontinuity, sensitivity or positively expansivity.

\section{Classification of Sensitive Elementary Cellular Automata}
\label{sec6}
\begin{figure}[ht]
\centering
\subfigure[\scriptsize{Rule 170}]{
\includegraphics[width=0.22\textwidth]{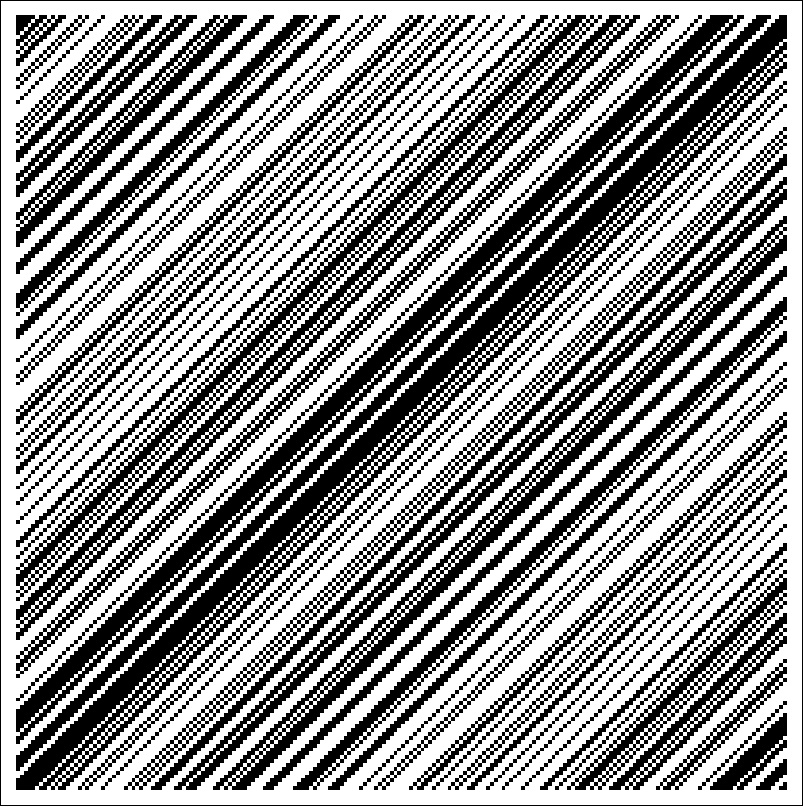}
\label{fig:subfig1}
}
\subfigure[\scriptsize{Rule 90}]{
\includegraphics[width=0.22\textwidth]{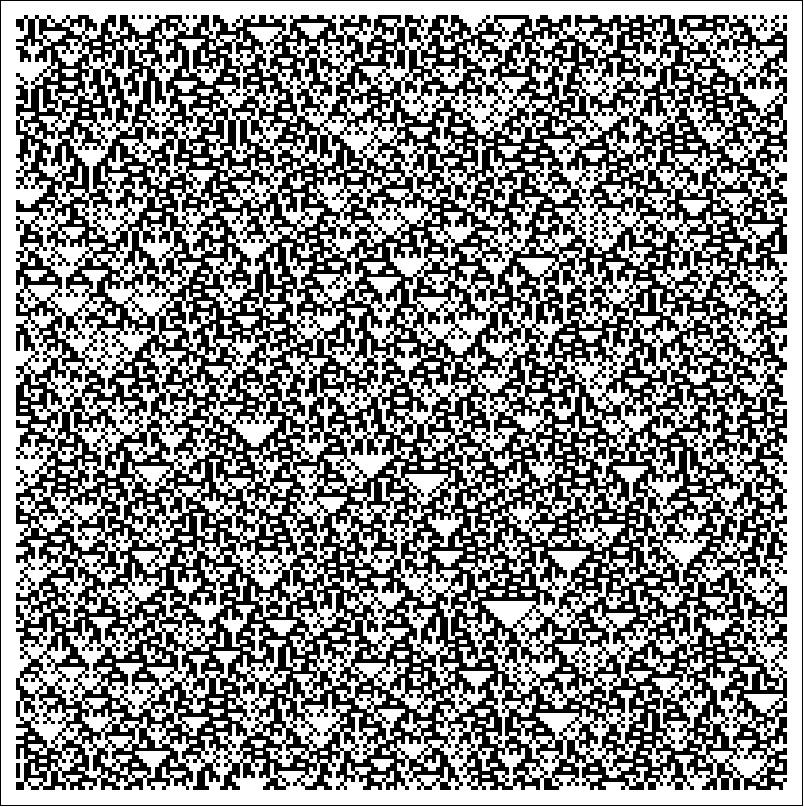}
\label{fig:subfig2}
}
\caption{\scriptsize{Space-time patterns of two chaotic ECA rules (rule 170 (a) and rule 90 (b)). The eventually weakly periodic ECA rule 170 simply shifts the values of cells and exhibits a ``travelling wave''. Finite arrays of 200 cells with periodic boundary conditions were used; black dots code state 1, white dots state 0. Time runs from top to bottom.}}
\label{fig_eca_weakperiodic}
\end{figure}

Since the beginning of CA research, the classification of the degree of ``complexity'' seen in CA behaviour has been a main research focus. It is intuitively clear that the sensitivity property is a source of the apparent ``complexity'' of ECA behaviour. Among the sensitive ECA rules, we find, however, rules that show in their space-time dynamics ``travelling waves'' patterns (Fig. \ref{fig_eca_weakperiodic}). These non-complex shift-dynamics patterns are from eventually weakly periodic ECA defined as follows.

A configuration $x$ is called \textit{weakly periodic}, if there is $q \in \mathbb{Z}$ and $p > 0$ such that $F^{p}\sigma^{q}(x)=x$.\cite{kurka2008topological} We define a configuration $x$ as \textit{eventually weakly periodic} if there is $q \in \mathbb{Z}$ and $n,p > 0$ such that $F^{n+p}\sigma^{q}(x)=F^{n}(x)$. We call an ECA \textit{eventually weakly periodic}, if the ECA is not eventually periodic, but for all configurations $x$ eventually weakly periodic.

\begin{proposition}
The following sensitive ECA rules are eventually weakly periodic:\\
2, 3, 10, 15, 24, 34, 38, 42, 46, 138, 170.
\label{prop21}
\end{proposition}
\begin{proof}
The general proof follows the argument exhibited for a specific ECA rule as follows.

Employing the algebraic expression for ECA rule 10, it can easily be shown that
$F^{2}\sigma^{-1}(x)=F^{}(x)$ for all configurations $x$.\\
Hence, ECA rule 10 is eventually weakly periodic with $n=1, p=1$ and $q=-1$.
\end{proof}
The classification of eventually weakly periodic ECA maps is not complete yet. There might be sensitive ECA which are eventually weakly periodic, but with such large $n$ or $p$ that prevents calculating the forward orbits as easily as in the proof of Proposition \ref{prop21}.

Surprisingly, some of the eventually weakly periodic ECA are also chaotic (but not positively expansive), while others are sensitive, but not chaotic. For this statement, we adhere to the standard definition of (topological) chaos given by Devaney \cite{devaney2003introduction}. A map $F: X \rightarrow X$ is chaotic, if $F$ is sensitive, transitive and if the set of periodic points of $F$ is dense in $X$. The class of chaotic ECA has already been determined by Cattaneo et al.\cite{cattaneo2000investigating}; for the sake of completeness, we rederive the result below. 

First, we shall study the surjectivity property shared by some ECA maps $F$. For sensitive ECA $F$, surjectivity is already sufficient to establish the transitivity of $F$ and the density of periodic points in $X$ under $F$, so that the chaoticity of $F$ is implied.

A CA is surjective if and only if it has no \textit{Garden-of-Eden} configurations, that is configurations which have no pre-image. A necessary (but not sufficient) condition for surjectivity is that the local rule is balanced.\cite{kurka2003topological} For ECA rules this means that the local rule table contains 4 zeros and 4 ones. Further, any permutive CA is surjective.\cite{kurka2003topological}

\begin{proposition}
The following ECA rules are surjective:\\
15, 30, 45, 51, 60, 90, 105, 106, 150, 154, 170, 204.
\label{prop23}
\end{proposition}
\begin{proof}
Apart from rule 51 and rule 204, the above listed rules are permutive, hence surjective. Rule 51 and rule 204 are surjective, because they are, trivially, bijective.\\ 
For the ECA rules that are not listed, but satisfy the balance condition, it can be shown that they possess Garden-of-Eden configurations. For example, ECA rule 184 satisfies the balance condition, nevertheless it is not surjective, because any configuration containing pattern $(1100)$ is a Garden-of-Eden as can easily be shown.
\end{proof}

Next, we show that for ECA transitivity is equivalent to permutivity. An one-dimensional CA $F$ is transitive if for any nonempty open sets, $U,V \subseteq X$ there exists $n>0$ with $F^n(U) \cap V \neq \emptyset$.
\begin{proposition}
A ECA is transitive if and only if it is permutive.
\label{prop16}
\end{proposition}
\begin{proof}
Transitivity of one-dimensional CA implies its surjectivity and sensitivity \cite{kurka2003topological}. From Proposition \ref{prop6} and \ref{prop23} and the definition of permutivity, we gain that ECA that are surjective and sensitive are permutive.\\
Conversely, permutive ECA are surjective \cite{kurka2003topological}. From the surjective ECA that are sensitive (the surjective ECA rules 51, 204 are not sensitive, hence not transitive), the positively expansive ECA are permutive and transitive \cite{kurka2003topological}. The ECA rules 15 and 170 are also permutive and transitive. Rule 106 is permutive and has been shown transitive \cite{kurka2003topological}. Proofs of the transitivity of the remaining permutive and sensitive rules 30, 45, 60, 154 can be similarly constructed.
\end{proof}

\begin{corollary}
A ECA map is transitive if and only if it is surjective and sensitive.
\label{cor23}
\end{corollary}
\begin{proof}
Follows from Proposition \ref{prop6}, \ref{prop23} and \ref{prop16}.
\end{proof}

Next, we show that for ECA surjectivity implies that the set of periodic points of $F$ is dense in $X$.
\begin{proposition}
Surjective ECA have a dense set of periodic points in $X$.
\label{prop25}
\end{proposition}
\begin{proof}
Surjective ECA are either almost equicontinuous or sensitive. Almost equicontinuous one-dimensional CA that are surjective have a dense set of periodic points \cite{blanchard2000some}. The sensitive ECA that are surjective are permutive and permutive one-dimensional CA are known to have a dense set of periodic points (through the property of closingness \cite{kurka2003topological}).
\end{proof}
While for general one-dimensional CA it is still an important open question whether surjectivity implies a dense set of periodic points, for ECA, transitivity or permutivity implies chaos.

\begin{corollary}
The following ECA rules are chaotic in the sense of Devaney:\\
15, 30, 45, 60, 90, 105, 106, 150, 154, 170.
\label{cor25}
\end{corollary}

The distinction between the chaotic and non-chaotic ECA is not necessarily seen in the space-time patterns. The eventually weakly periodic ECA that are chaotic and the eventually weakly periodic ECA that are sensitive but not chaotic both show similar ``travelling wave'' patterns. The difference between the chaotic ECA and the sensitive but not chaotic ECA is not in the space-time patterns they generate, but in how they react to perturbations.

\begin{figure*}
  \begin{center}
    \begin{tabular}{c}
\includegraphics[width=0.85\textwidth]{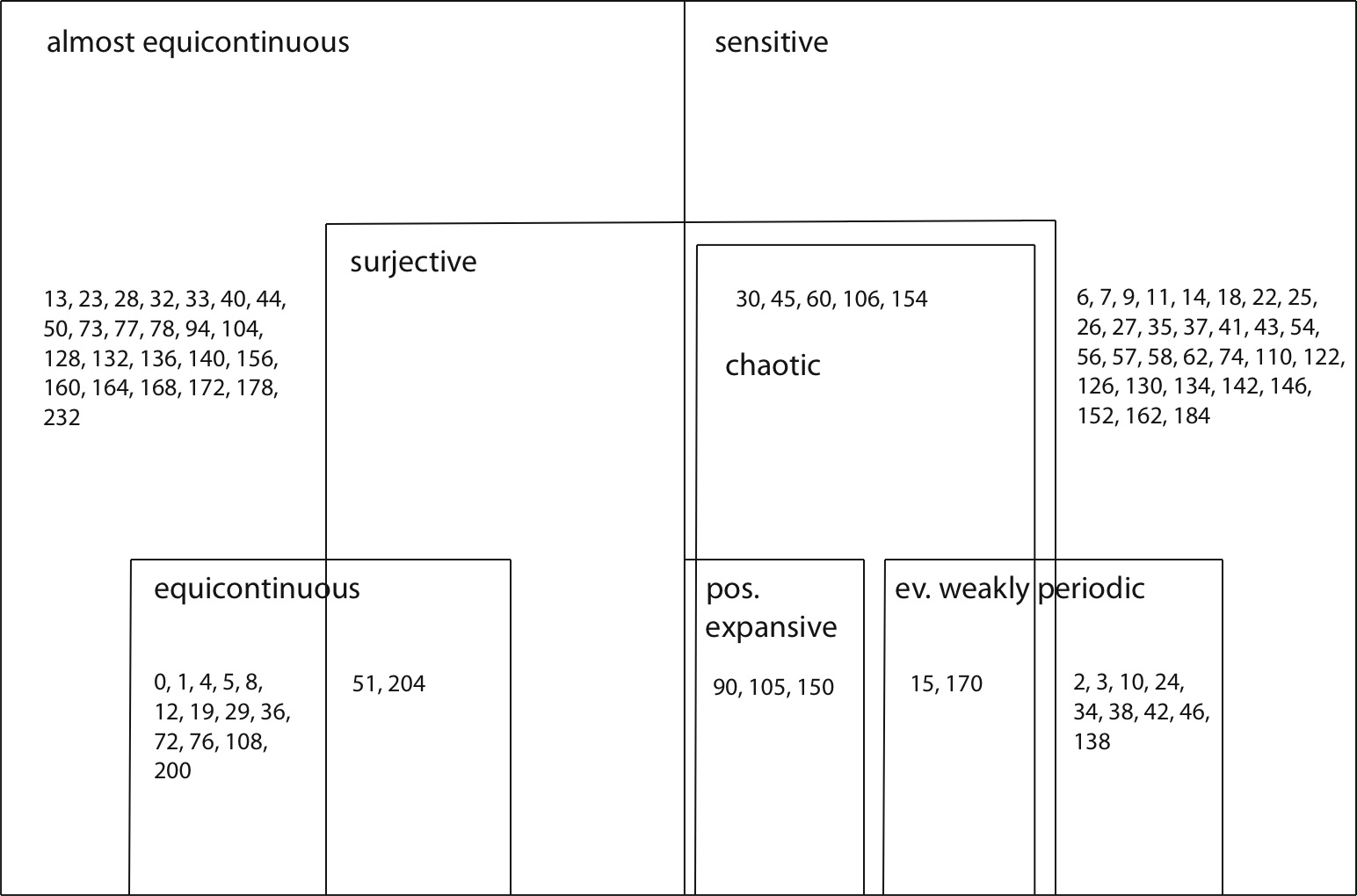}
    \end{tabular}
    \caption[]{\footnotesize Classification diagram for the elementary cellular automata (ECA). The chaotic ECA are inside the double-framed box. The class of the sensitive and eventually weakly periodic ECA is not complete.} 
 \label{fig2} 
 \end{center}
\end{figure*}

While the eventually weakly periodic ECA show too simple behaviour to be called ``complex'', chaotic ECA are in a sense ``too complex'': their mixing properties do no allow for the memory capacities apparently needed for ``complex'' behaviour. In the following final section we will expand on this observation. Figure \ref{fig2} summarises the results of our analysis.

\section{Discussion}
\label{sec7}
The results of this paper show that one can classify the dynamic behaviour of every elementary cellular automaton (ECA) in terms of the standard notions of dynamical system theory, that is, according to the classification proposed by Gilman \cite{gilman1987classes} and Kurka \cite{kurka2001languages}. We also determined which ECA are chaotic in the sense of Devaney, rederiving a result by Cattaneo et al. \cite{cattaneo2000investigating} This gives a fairly complete picture of the dynamical system properties of ECA in the standard topology, as summarised in Fig. \ref{fig2}. The topological dynamics approach to CA thus delivers a relevant and coherent account of the dynamical behaviour of ECA.

In the light of our results, the class of ``complex'' ECA can be characterised as those ECA that are sensitive, but not surjective, and not eventually weakly periodic. This class corresponds well to what one would intuitively regard as ``complex'', given the space-time patterns of ECA. In particular, the ECA rules of Wolfram's class (W4) seem to fall into this class. 

Among the ECA rules, a few deserve special interest from a computational point of view. The most prominent example is ECA rule 110 which has been shown to be computationally universal.\cite{cook2004universality} Based on our results we conjecture that sensitivity is a necessary condition of computational universality. In contrast, Wolfram conjectured that, for example, ECA rule 73, which is not sensitive, may be computationally universal.\cite{wolfram2002new} This difference is due to the fact that our results hold generally for ECA without any restrictions on the initial conditions, whereas Wolfram considers specific sets of initial configurations on which the rule acts. On such a restricted set of configurations, ECA rule 73 might indeed be sensitive. 

If a CA is sensitive, then its dynamics defies numerical computation for practical purposes, because a finite precision computation of an orbit may result in a completely different orbit than the real orbit. Hence, while sensitivity seems inherent to the in principle computationally most powerful rules, as e.g. rule 110, their limited robustness to small changes in the initial conditions may impair their practical usage in a physical or biological system: Even a single bit-flip in the input of a sensitive ECA may completely change the computed output.

Among the many questions left open, a natural extension of our study would consist in giving a complete characterisations in the topological dynamics sense for more general CA than ECA. Examples by Cattaneo et al. \cite{cattaneo2000investigating}, however, show that the approach taken here to establish chaoticity can already fail in slightly more general settings. In the general case, long-term properties of CA and hence classification schemes based on these properties are typically undecidable. It would therefore be useful to pinpoint where exactly undecidability enters. 

Establishing a verifiable notion of computational universality in the Turing-machine sense in terms of necessary and sufficient conditions related to the dynamic behaviour of the underlying system would greatly advance our understanding of the relation between computational and dynamic properties of physical and biological systems. Part of the problem to clarify this relation is that there is no unanimous accepted definition of computational universality for computational systems such as CA (see e.g. the discussion by Ollinger \cite{ollinger2008universalities} and Delvenne et al.\cite{delvenne2005computational} Delvenne et al. also prove necessary conditions for a symbolic system to be universal, according to their definition of universality, and demonstrate the existence of a universal and chaotic system on the Cantor space.). To different definitions of universality, there might thus correspond different topological dynamics properties. Despite this fact, we conjecture that for ECA sensitivity and non-surjectivity are necessary conditions of universality. This conjecture is in accordance with the intuitive idea that systems at the ``edge of chaos'', i.e. systems with neither too simple nor chaotic dynamical behaviour, are the computationally relevant systems for biology. Such intermittent systems have, moreover, been characterised as having the largest complexity in the sense that their behaviour is the hardest to predict.\cite{stoop2004complexity} If computation is measured as a reduction of complexity \cite{stoop2004natural}, the intermittent systems may then be said to provide the complexity needed for efficient computations. 

The extension of the results and observations from ECA to general one-dimensional CA or higher-dimensional CA is thus not without problems. Being much more tractable, ECA provide an important benchmark to test ideas on universality, the ``edge of chaos'' hypothesis and, generally, on how ``computation'' occurs in nature.

\section*{Acknowledgments}
The authors wish to thank Jarkko Kari for helpful comments and advice.\\ This work was partly supported by ETH Research Grant TH-04 07-2 and the cogito foundation.


\end{document}